\documentclass[genTeX]{nrc1}

\usepackage{cite} 
\usepackage{amsmath} 
\usepackage{amsfonts}
\usepackage{bm} 
\usepackage{bbm} 
\usepackage{epsf}
\usepackage{times}
\usepackage{nicefrac}
\usepackage{graphicx}

\def\buildrel#1\under#2{\mathrel{\mathop{\kern0pt #2}\limits_{#1}}}

\def\bm#1{\mbox{\boldmath$#1$}}

\def\corresponds{\sim}
\def\succsim{\succ\kern-.9em_\sim\kern.3em}
\def\precsim{\prec\kern-1em_\sim\kern.3em}
\def\slantfrac#1#2{\kern1em^{#1}\kern-.3em/\kern-.1em_{#2}}
\def\lfrac#1#2{{}^{#1\!}\kern-.0em/_{#2}}

\journal{Canadian Journal of Physics}

\volyear{1}[2004]{2004}

\begin{document}

\title{Some Recent Advances in Bound--State Quantum Electrodynamics}

\author{U.~D.~Jentschura and J.~Evers}

\address{Max--Planck--Institut f\"ur Kernphysik,
Saupfercheckweg 1, 69117 Heidelberg, Germany}

\maketitle

\shortauthor{Jentschura and Evers}

\begin{abstract}
We discuss recent progress in various problems
related to bound-state quantum electrodynamics:
the bound-electron $g$ factor,
two-loop self-energy corrections 
and the laser-dressed Lamb shift.
The progress relies on various advances 
in the bound-state formalism, including 
ideas inspired by effective field theories
such as Nonrelativistic Quantum Electrodynamics.
Radiative corrections in dynamical processes
represent a promising field for further 
investigations.
\end{abstract}

\keywords{Calculations and mathematical techniques in atomic and
molecular physics, \\
quantum electrodynamics -- specific calculations;\\
{\it PACS numbers:} 31.15.-p, 12.20.Ds}

%
%
\section{Introduction}

This is a brief summary of a number of recent advances
in our understanding of bound-state quantum electrodynamic
(QED) effects. The topics are (i) two-loop corrections
to the bound-electron $g$ factor, (ii) higher-order 
two-loop corrections to the self-energy of a bound electron,
and (iii) the laser-dressed Lamb shift. 

The first two of these rather diverse topics are 
related to two-loop effects. The investigation 
of these is simplified considerably by the use of 
effective field-theory techniques inspired
by Nonrelativistic QED (NRQED)~\cite{CaLe1986,Pa1993,PiSo1998}.
The Wilson coefficients multiplying the effective 
operators in the NRQED Lagrangian are matched against
those of the full relativistic theory, providing a simplified 
framework for the calculation of bound-state effects.
Scale-separation parameters such as the photon mass $\mu$
are cancelled at the end of the calculation.
The analysis of higher-order corrections to the $g$ factor 
of the bound electron is simplified further by 
a transformation to the length gauge, which results in 
a lesser number of terms to be considered than would be 
necessary in the velocity gauge. This fact has inspired 
the development of Long-wavelength QED (LWQED)~\cite{Pa2004},
a theory which is obtained after Power--Zienau and
Foldy--Wouthuysen transformations of the first-quantized 
Lagrangian; the second quantization is carried out by formulating
the path integral. Consequently, an improved understanding
and a tremendous simplification results for the 
calculation of a number of QED corrections for bound states,
such as the $g$ factor and higher-order corrections to the 
self-energy.

A further field of recent studies has been concerned with 
the interaction of a laser-dressed bound electron with the radiation 
field~\cite{JeEvHaKe2003,JeKe2004aop,EvJeKe2004,JeEvKe2004}. 
This process entails corrections which can only be understood
if the analysis is carried out right from the start
in the framework of the laser-dressed states, which are 
the eigenstates of the quantized atom-laser Hamiltonian
in the rotating-wave approximation~\cite{CT1975}.

%
%
\section{Bound-electron $g$ factor}

In this section, we briefly summarize the 
results of a recent investigation~\cite{PaJeYe2004}
of the bound-electron $g$ factor, which is based on 
NRQED. The central result of this investigation is the 
following semianalytic expansion 
in powers of $Z\alpha$ and $\ln(Z\alpha)$
for the bound-electron $g$ factor ($n{\rm S}$ state)
in the non-recoil limit, which is 
the limit of an infinite nuclear mass:
\begin{align}
\label{gbound}
g(n{\rm S}) & = 
\underbrace{2 - \frac{2\, (Z\alpha)^2}{3\,n^2} + 
\frac{(Z\alpha)^4}{n^3} \, \left( \frac{1}{2 n} - \frac23 \right) + 
{\cal O}{(Z\alpha)^6}}_{\mbox{Breit (1928),
Dirac theory}}
\nonumber\\[4ex]
&  
+ \underbrace{{\frac{\alpha}{\pi}}\, 
\left\{ {2 \times \frac{1}{2}}\, {\bigg(}
1 + \frac{(Z\alpha)^2}{6 n^2} {\bigg)}
+ \frac{(Z\alpha)^4}{n^3} {\bigg\{}
a_{41}\, {\ln}[(Z\alpha)^{-2}]  + 
a_{40} {\bigg\}}
+ {\cal O}(Z\alpha)^5 \right\}}_{\mbox{one-loop correction}}
\nonumber\\[4ex]
& 
+ \underbrace{{ \left(\frac{\alpha}{\pi}\right)^2}\,  
\left\{ {-0.656958} \,  {\bigg(} 1 + 
\frac{(Z\alpha)^2}{6 n^2} {\bigg)}  + 
\frac{(Z\alpha)^4}{n^3} {\bigg\{}
b_{41}\,
{\ln}[(Z\alpha)^{-2}] + 
b_{40} {\bigg\}}
+ {\cal O}(Z\alpha)^5 \right\}}_{\mbox{two-loop correction}}
\nonumber\\[4ex]
& + {\cal O}(\alpha^3)\,.
\end{align}
This expansion is valid through the order
of two loops (terms of order $\alpha^3$ are 
neglected). The notation is in part inspired by the usual conventions
for Lamb-shift coefficients: the (lower case) $a$ terms denote the 
one-loop effects, with $a_{kj}$ denoting the 
coefficient of a term proportional to 
$\alpha\,(Z\alpha)^k\,{\ln}^j[(Z\alpha)^{-2}]$.
The $b$ terms denote the two-loop corrections,
with $b_{kj}$ multiplying 
a term proportional to 
$\alpha^2\,(Z\alpha)^k\,{\ln}^j[(Z\alpha)^{-2}]$.
In~\cite{PaJeYe2004}, complete results are derived for the 
coefficients $a_{41}$, $a_{40}$ and $b_{41}$.

In general, the expression corresponding to (\ref{gbound}) 
for a free electron is obtained by letting the 
parameter $Z\alpha \to 0$ in every term of the loop
expansion (expansion in powers of $\alpha$).
In this limit, the known free-electron two-loop result is 
recovered~\cite{KaKr1950,So1957,So1958,Pe1957helv,Re1972a,Re1972b,Ad1989}.

\begin{figure} 
\begin{center}
\begin{minipage}{10cm}
\begin{center}
\includegraphics[width=0.8\linewidth]{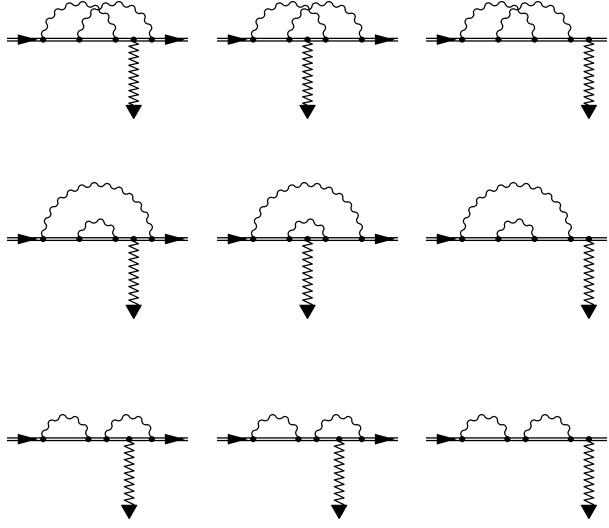}
\end{center}
\caption{\label{fig1} Feynman diagrams 
for the two-loop self-energy corrections
to the bound-electron $g$-factor.}
\end{minipage}
\end{center}
\end{figure}

Up to the relative order $(Z\alpha)^2$, the free-electron 
contribution in one-, two-, and higher-loop
order is multiplied by a relative factor
\begin{equation}
\label{Zalpha2}
1 + (Z\alpha)^2 \, a_{20} = 
1 + (Z\alpha)^2 \, b_{20} = 1 + \frac{(Z\alpha)^2}{6 n^2}\,.
\end{equation}
This result consequently holds for the three-loop and the 
four-loop term not shown in Eq.~(\ref{gbound}). 
The applicability of the relative factor (\ref{Zalpha2})
to the two-loop term, valid through $(Z\alpha)^2$,
had been stressed previously in~\cite{Ka2001proc}.
The result in Eq.~(\ref{Zalpha2}) had been obtained originally 
in~\cite{Gr1970prl,Gr1970pra,GrHe1971,Fa1970plb,Fa1970nc}
(for the 1S state).
As is evident from Eq.~(\ref{gbound}), 
the correction of relative order $(Z\alpha)^2$
is different on the level of the tree-level diagrams
and reads $g(n{\rm S}) \sim 2 \times [1 - (Z\alpha)^2/(3 n^2)]$.

Explicit results for the coefficients in (\ref{gbound}),
restricted to the one-loop self-energy, read~\cite{PaJeYe2004}
\begin{subequations}
\label{a4}
\begin{align}
\label{a41}
a_{41}(n{\rm S}) &= \frac{32}{9}\,, \\[2ex]
\label{a40}
a_{40}(n{\rm S}) &= \frac{73}{54} 
- \frac{5}{24 n}
- \frac{8}{9} \, \ln k_0(n{\rm S}) 
- \frac{8}{3} \, \ln k_3(n{\rm S})\,.
\end{align}
\end{subequations}
Here, $\ln k_0(nS)$ is the Bethe logarithm for an $nS$ state,
and $\ln k_3(nS)$ is a generalization of the Bethe logarithm
to a perturbative potential of the form $1/r^3$.
Vacuum polarization adds a further $n$-independent
contribution of $(-16/15)$ to $a_{40}$~\cite{Ka2000}.
The Bethe logarithms for 1S and 2S~\cite{DrSw1990}
read
\begin{subequations}
\label{lnk0}
\begin{eqnarray}
\label{lnk01S}
\ln k_0(1{S}) &=& 2.984~128~555\,,\\[1ex]
\label{lnk02S}
\ln k_0(2{S}) &=& 2.811~769~893\,,
\end{eqnarray}
\end{subequations}
and the corresponding values for $\ln k_3$ read~\cite{PaJeYe2004}
\begin{subequations}
\label{lnk3}
\begin{eqnarray}
\label{lnk31S}
\ln k_3(1{S}) &=& 3.272~806~545\,,\\[1ex]
\label{lnk32S}
\ln k_3(2{S}) &=& 3.546~018~666\,.
\end{eqnarray}
\end{subequations}
The quantity $\ln k_3(n{\rm S})$ is defined as,
\begin{equation}
\int_0^\epsilon {\rm d} k\, k^2\,
\langle\phi|\bm{r}\,\frac{1}{E-H_S-k}\,\frac{1}{r^3}\,
\frac{1}{E-H_S-k}\,\bm{r}\,|\phi\rangle
= -4\,\frac{(Z\alpha)^3}{n^3} \,
\left[\ln \frac{\mu}{(Z\alpha)^2}+\frac56-\ln k_3\right]
\end{equation}
where the ultraviolet cutoff $\epsilon$ is to be understood 
in the sense of~\cite{Pa1993}, and the matching of the noncovariant 
cutoff $\epsilon$ to the covariant photon mass $\mu$ 
is given as (see~\cite{ItZu1980}, pp.~361--362)
\begin{equation}
\ln \left( \frac{2\epsilon}{m} \right) \to 
\ln \left(\frac{\mu}{m}\right) + \frac{5}{6}\,.
\end{equation}
However, this replacement is not unique and the constant
term depends on the actual form of the integrand. A different
replacement has to be used for some of the low-energy photon
corrections to the $g$ factor~\cite{PaJeYe2004}.

The results for the two-loop coefficients read
\begin{subequations}
\label{b4}
\begin{align}
\label{b41}
b_{41}(n{\rm S}) &= \frac{56}{9}\,, \\[2ex]
\label{b40}
b_{40}(1{\rm S}) &= -18.5(5.5)\,.
\end{align}
\end{subequations}
Here, the result for $b_{40}$ is an estimate based
on an explicit calculation of a large contribution due to 
low-energy virtual photons, and an estimate
of the remaining, unknown contribution due to high-energy virtual
photons. The dominant logarithmic two-loop
term $b_{41}$ is caused
exclusively by the two-loop self-energy diagrams in Fig.~\ref{fig1}
alone. The other two-loop diagrams, 
which include closed fermion loops,
can be found in Fig.~21 of~\cite{Be2000}.
The logarithmic term $b_{41}$ is, however,
exclusively related to the gauge-invariant 
subset displayed in Fig.~\ref{fig1}.

The newly calculated $a_{41}$, $a_{40}$ and $b_{41}$
are bound-state corrections to the electron $g$-factor
of order $(\alpha/\pi)\, (Z\alpha)^4$ and
$(\alpha/\pi)^2\, (Z\alpha)^4$, multiplied by 
logarithmic terms. These corrections
are (at $Z=1$) formally of order $\alpha^5$ and $\alpha^6$
and therefore of the same order of magnitude as
the tenth- and twelfth-order corrections to the
free-electron anomaly, which barely are of experimental
or theoretical significance at the current level
of accuracy. One may therefore ask why these binding
corrections are of any phenomenological significance.
The reason is that at somewhat higher $Z$,
the situation changes drastically, due to $Z^4$
scaling of the binding corrections. In addition,
due to numerically large coefficients and logarithmic
factors, the ``hierarchy'' of the corrections changes drastically.
Roughly, one may say that at $Z=1$, the bound-electron
anomalous magnetic moment is approximately independent
of binding corrections of order  $(\alpha/\pi)\, (Z\alpha)^4$
and higher, whereas for higher $Z$, the situation is reversed,
and the binding corrections to the one- and two-loop
contributions are numerically much more 
significant than the higher-loop 
free-electron corrections. This ``transition
from free to bound-state quantum electrodynamics'' as a
function of $Z$ is a somewhat
peculiar feature of the bound-electron $g$-factor.

For, example, we consider the ratio
\begin{equation}
\label{r1def}
r_1(Z) = \frac{\displaystyle 
\left(\frac{\alpha}{\pi}\right)\,(Z\alpha)^4\,\ln[(Z\alpha)^{-2}]}
{\displaystyle \left(\alpha/\pi\right)^4}\,,
\end{equation}
which gives an order-of-magnitude estimate for the the ratio of the
one-loop self-energy binding correction to the
eighth-order anomalous magnetic moment of the free electron.
We have
\begin{equation}
\label{r1num}
r_1(Z=1) \approx 2\,, \qquad 
r_1(Z=10) \approx 10^4 \gg 1\,. \qquad 
\end{equation}
For the two-loop logarithmic binding correction, we
have
\begin{equation}
\label{r2def}
r_2(Z) = \frac{\displaystyle 
\left(\frac{\alpha}{\pi}\right)^2\,(Z\alpha)^4\,\ln[(Z\alpha)^{-2}]}
{\displaystyle \left(\alpha/\pi\right)^4}
\end{equation}
and consequently
\begin{equation}
\label{r2num}
r_2(Z=1) \approx 0.2 < 1\,, \qquad 
r_2(Z=10) \approx 2 \times 10^3 \gg 1 \,. \qquad 
\end{equation}
As is evident from these considerations, the 
$(Z\alpha)^4$ one-loop and two-loop binding corrections are 
roughly of the same order of magnitude as the 
highly problematic four-loop 
corrections~\cite{KiLi1990,Ki1995ieee,HuKi1999} for the free 
electron. However, the situation changes drastically even at
very moderate nuclear charge numbers, and the binding 
corrections to the one-loop and two-loop contributions
become dominant over the higher-loop effects. 

The NRQED one-loop calculation~\cite{PaJeYe2004} is divided into 
three parts, the first of which entails fully relativistic 
form-factor corrections including lower-order terms, 
the second of which corresponds to a spin-dependent 
scattering amplitude, and the third of which is a 
low-energy Bethe-logarithm type correction that contains
$\ln k_0$ and $\ln k_3$. The one-loop
correction in Eq.~(\ref{gbound}) can therefore be 
written in a natural way as 
$ \delta g^{(1)} = g_1^{(1)} + g_2^{(1)} + g_3^{(1)}$,
where

\begin{subequations}
\label{g1}
\begin{align}
\label{g11}
g_1^{(1)} &= \frac{\alpha}{\pi}\biggl[ 1+\frac{(Z\alpha)^2}{6\,n^2} -
\frac{(Z\alpha)^4}{n^3}\,\left(\frac76
  +\frac5{24n} +\frac{16}{3}\, \ln \mu \right) \biggr] \,, \\[2ex]
\label{g21}
g_2^{(1)} &= \frac{\alpha}{\pi}\,\frac{(Z\alpha)^4}{n^3}\,
\Bigl(4+\frac{16}{9}\,\ln\mu\Bigr)\,, \\[2ex]
\label{g31}
g_3^{(1)} &= \frac{\alpha}{\pi}\,\frac{(Z\alpha)^4}{n^3}\,\frac{32}{9}\,
\biggl[\ln\frac{\mu}{(Z\alpha)^2} -\frac{5}{12} -\frac{\ln k_0}{4}
-\frac{3}{4}\,\ln k_3 \biggr]\,.
\end{align}
\end{subequations}

The new contribution of order $(Z\alpha)^4$ can be
compared with the numerical results for the self-energy 
correction~\cite{YeInSh2002,YeInSh2004} complete to all orders in $Z\alpha$. 
Assuming correctness of the logarithmic term in Eq.~(\ref{g1}), 
a fit to numerical data yields 
$a_{40}^{(1)}({\rm 1S}) = -10.2(1)$ and 
$a_{40}^{(1)}({\rm 2S}) = -10.6(1.2)$ for the constant term, in excellent
agreement with the analytic results which read
$a_{40}^{(1)}({\rm 1S}) = -10.236~524~318(1)$ 
and $a_{40}^{(1)}({\rm 2S}) = -10.707~715~607(1)$ . 

Having verified the consistency 
of the analytic~\cite{PaJeYe2004}
and numerical results~\cite{YeInSh2002,YeInSh2004}, an interpolation 
procedure~\cite{IvKa2001proc} may now be used to extract 
a more accurate theoretical prediction 
at low and intermediate nuclear charge numbers, 
if combined with numerical results at
higher $Z$~\cite{YeInSh2002}. Thus, the 
results in Eqs.~(\ref{a4}) and (\ref{b4}) 
may be used in order to infer improved theoretical 
predictions for the bound-electron $g$ factor, 
notably in the experimentally important 
special cases of hydrogenlike carbon~\cite{HaEtAl2000prl}
and oxygen~\cite{VeEtAl2004}. Alternatively, the 
improved status of the theory may be used in order to 
infer a more accurate value of the 
electron mass. Specifically, the value from 
the carbon measurement~\cite{HaEtAl2000prl},
using the new theory, reads
\begin{equation}
m(^{12}{\rm C}^{5+}) = 0.000\,548\,579\,909\,41\, (29)(3)\ {\rm u} \,. 
\end{equation}
The first error comes from the 
experiment~\cite{HaEtAl2000prl}, and the 
second error corresponds to the theoretical uncertainty. 
The conclusion is that a further 
improvement of the experiment could lead to a much 
better determination of the electron 
mass; the new theory provides room 
for at least an improved determination by one order of magnitude.

For the calculation of yet higher-order binding corrections
to the one-loop and two-loop contributions,
a detailed understanding of the two-loop form-factors,
including their slopes, is required. 
The most recent calculations of these effects,
in both dimensional and photon-mass regularizations,
can be found in~\cite{BoMaRe2003,MaRe2003,BoMaRe2004}.

\begin{figure}[htb]
\begin{center}
\begin{minipage}{14cm}
\begin{center}
\includegraphics[width=1.0\linewidth]{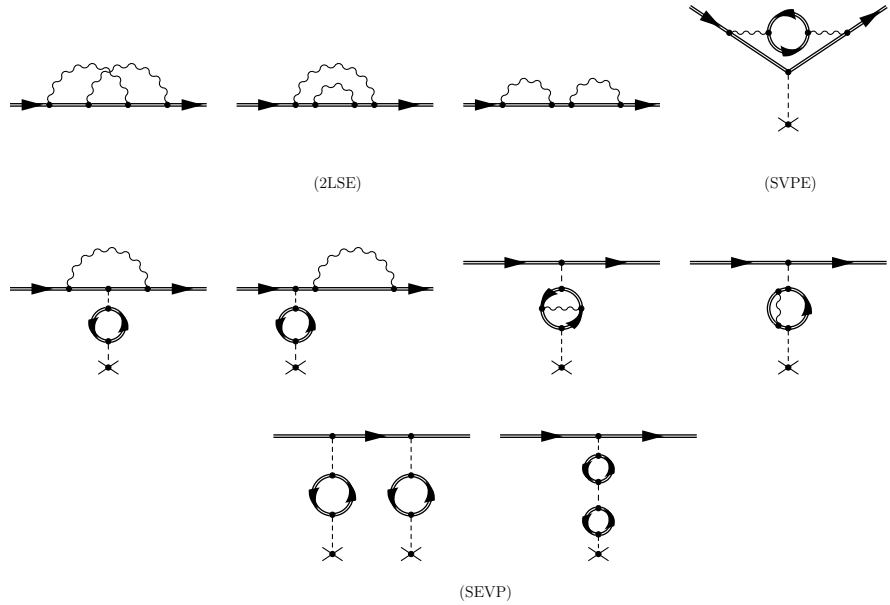}
\end{center}
\caption{\label{fig2} The two-loop corrections to the 
Lamb shift in hydrogenlike systems fall naturally 
into three gauge-invariant subsets, which are the 
pure two-loop self-energy terms (2LSE), the 
vacuum-polarization correction to the virtual-photon  
line in the self-energy (SVPE), and the 
self-energy vacuum-polarization and pure 
two-loop vacuum-polarization corrections (SEVP).
The two-loop Bethe logarithm is a numerically 
large correction to the nonlogarithmic term of order
$\alpha^2 (Z\alpha)^6$ and it is exclusively related 
to the 2LSE subset; however, for a complete result
in this order, contributions from the other gauge-invariant subsets 
will have to be considered as well.}
\end{minipage}
\end{center}
\end{figure}

%
%
\section{Two-loop Bethe logarithms}

As is well known~\cite{ApBr1970,Ka1993log,Pa1994prl,%
EiSh1995,Ka1996,Ka1996b,Ka1997,Pa2001}, 
the two-loop Lamb shift $\Delta E^{(2)}$, 
in the limit of an infinite nuclear mass, may be written as
\begin{eqnarray}
\Delta E^{(2)} =
\left(\frac{\alpha}{\pi}\right)^2 \, 
\frac{(Z\alpha)^4 \, m_{\rm e} \,c^2}{n^3} \,
H(Z\alpha)\,.
\end{eqnarray}
For S states, the dimensionless function $H(Z\alpha)$,
has a semianalytic expansion of the form
\begin{align}
H(Z\alpha) = &  B_{40} + (Z\alpha) \, B_{50}
\nonumber\\[2ex]
& + (Z\alpha)^2 \, \left[ B_{63} \,
\ln^3[(Z\alpha)^{-2}] + B_{62} \, \ln^2[(Z\alpha)^{-2}]
+ B_{61} \, \ln[(Z\alpha)^{-2}] + B_{60} \right] \,,
\end{align}
where we ignore higher-order terms, and upper case is used 
for the $B_{ij}$ coefficients that multiply terms 
of order $\alpha^2\,(Z\alpha)^i\,\ln^j[(Z\alpha)^{-2}] \, m_{\rm e} \,c^2$.
The coefficients, restricted to the 
two-photon self-energy diagrams (Fig.~\ref{fig2}),
read as follows
\begin{subequations}
\begin{eqnarray}
B^{\rm (2LSE)}_{63}(n{\rm S}) &=&
-\frac{8}{27} = -0.296296\,, \\[2ex]
B^{\rm (2LSE)}_{62}(1{\rm S}) &=& 
\frac{16}{27} - \frac{16}{9}\, \ln(2) 
= -0.639\,669\,, \\[2ex]
B^{\rm (2LSE)}_{62}(n{\rm S}) &=& 
B^{\rm (2LSE)}_{62}(1{\rm S}) + 
\frac{16}{9} \left( \frac34 + \frac{1}{4 n^2} -
\frac1n - \ln(n) \! + \! \Psi(n) + C\right)\,.
\end{eqnarray}
\end{subequations}

The $n$-dependence of $B_{61}$ has been 
clarified in~\cite{Pa2001,Je2003jpa}
\begin{eqnarray}
B^{\rm (2LSE)}_{61}(n{\rm S}) &=& B^{\rm (2LSE)}_{61}(1{\rm S})
+ \frac43 \, \left[ N(n{\rm S}) - N(1{\rm S})\right] 
\nonumber\\[0.5ex]
& & + \! \left( \frac{80}{27} - \frac{32}{9} \, \ln (2)\right) 
\left(\frac34 + \frac{1}{4 n^2} - \frac1n - \ln(n) \! + \! \Psi(n) 
\! + \! C\right)\,, 
\end{eqnarray}
where $C=0.577216\dots$ is Euler's constant,
$\Psi(n)$ is the logarithmic derivative of the Gamma function, 
and $N(n{\rm S})$ is related to a correction to the
Bethe logarithm induced by a Dirac-delta potential.
Explicit values for $N(n{\rm S})$ can be found in~\cite{Je2003jpa}
($n = 1,\dots,8$).

\begin{figure}[htb]
\begin{center}
\begin{minipage}{8cm}
\begin{center}
\includegraphics[width=0.9\linewidth]{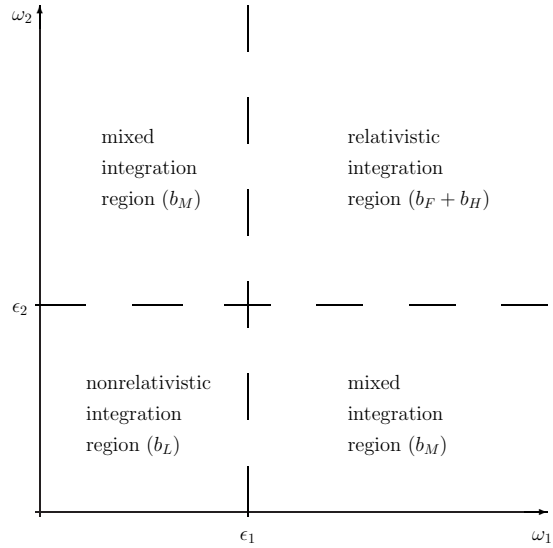}
\caption{\label{fig3}
The integration regions for the two virtual photons in the 
two-loop self-energy problem comprise a low-energy regime 
with two low-energy photons, which gives rise to $b_L$.
The middle-energy regions with one low-energy and a one 
high-energy photon give rise to $b_M$. The 
high-energy contribution $b_F + b_H$ is as yet unknown.}
\end{center}
\end{minipage}
\end{center}
\end{figure}

The $B_{60}$ coefficients are the sum of several 
contributions
\begin{subequations}
\begin{eqnarray}
\label{2lse}
B^{\rm (2LSE)}_{60}(n{\rm S}) &=&
b_L + b_M + \left\{ b_F + b_H + b_{\rm VP} \right\}\,, \\[2ex]
b_L &=& b_L(n{\rm S}) \; \corresponds \;
\mbox{two-loop Bethe logarithm, two soft photons}\,,
\\[2ex]
b_M &=& b_M(n{\rm S}) \; = \;
\frac{10}{9} \, N(n{\rm S})
\; \corresponds \; \mbox{one soft, one hard photon}\,,
\\[2ex]
b_F &\corresponds&
\mbox{soft electron momenta, two hard photons}\,,
\\[2ex]
b_H &\corresponds&
\mbox{hard electron momenta, two hard photons}\,,
\\[2ex]
b_{\rm VP} &\corresponds&
\mbox{vacuum-polarization corrections}\,.
\end{eqnarray}
\end{subequations}
Only the terms $b_L + b_M$ are currently 
known~\cite{PaJe2003,Je2004b60}~(see also Fig.~\ref{fig2}).
The contributions in curly brackets in 
Eq.~(\ref{2lse}) remain to be evaluated.
However, an estimate for the total value of $B_{60}$
may be obtained, 
\begin{eqnarray}
B_{60}(n{\rm S}) &=&
b_L + b_M + \left\{ b_F + b_H + b_{\rm VP} \right\}\,, \nonumber\\[2ex]
B_{60}(n{\rm S}) &\approx&
b_L + b_M \; \pm \; 15\,\%\,.
\end{eqnarray}
This estimate~\cite{PaJe2003,Je2004b60} is based on corresponding 
one-loop calculations,
where the low-energy virtual photons give the by far 
dominant contribution to the constant term~\cite{Pa1993}.
The results for the two-loop Bethe logarithms of S states 
read~\cite{PaJe2003,Je2004b60}
\begin{subequations}
\label{bL}
\begin{eqnarray}
b_L(1S) &=& -81.4(3)\,, \\[2ex]
b_L(2S) &=& -66.6(3)\,, \\[2ex]
b_L(3S) &=& -63.5(6)\,, \\[2ex]
b_L(4S) &=& -61.8(8)\,, \\[2ex]
b_L(5S) &=& -60.6(8)\,, \\[2ex]
b_L(6S) &=& -59.8(8)\,.
\end{eqnarray}
\end{subequations}
A few clarifying remark might be in order. The $B_{60}$ coefficient
multiplies a correction of order
$\alpha^2 (Z\alpha)^6$, which is effectively an order-$\alpha^8$ 
contribution to the energy levels of hydrogen ($Z=1$). 
In order to complete the calculation at this 
order of magnitude, it would also be necessary to 
consider the four-loop Dirac form-factor slope of the electron,
as well as the three-loop binding correction
of order $\alpha^3\,(Z\alpha)^5\,m_{\rm e}\,c^2$~\cite{EiSh2004}.
The three-loop slope has recently been evaluated 
in~\cite{MeRi2000}, completing the theory of 
energy levels in hydrogen up to the order of $\alpha^7$.

%
%
\section{Laser-dressed Lamb shift}

In the recent past, seminal advances have been obtained both 
in the techniques
of high-precision spectroscopy (e.g.,~\cite{ReEtAl2000}),
and in the coherent preparation 
and manipulation of media by external electromagnetic 
fields~\cite{ScZu1997,FiSw2005}.
Thus it is desirable to study the bound electrons 
interacting simultaneously both with the quantized 
radiation field and with an external driving 
field. An accurate theory of such systems,
including all dynamic effects, might eventually
open a possibility for a whole new class of high-precision
experiments, provided that technical problems related
to the required highly accurate intensity stabilization of the laser 
(and others) can be solved. 
Traditionally, radiative and relativistic corrections 
are treated with methods of QED, whereas studies 
related to the dynamical nature of the interaction of matter 
with driving laser fields are the domain of Quantum Optics (QO).
Obviously, a treatment of bound electrons in the presence
of both the radiation field and external driving fields requires
a combination of ideas from both subject areas:
While, {\em a priori}, the essential-state approximation
of QO~\cite{ScZu1997} is not sufficient to obtain the accuracy
of QED, a perturbative treatment of the interaction of the bound
electron with a strong external (laser) field as in QED is hopeless
because of the large coupling parameter.

In~\cite{JeEvHaKe2003,JeKe2004aop,EvJeKe2004}, 
an atom with two relevant energy levels driven by a strong
near-resonant monochromatic laser field is studied as the easiest
model system for the above problem. The incoherent
part of the resonance fluorescence spectrum emitted by this system
in QO is known as the Mollow spectrum, where the coupling strength
is characterized by the the Rabi frequency
$\Omega$ defined as ($\hbar = \epsilon_0 = c = 1$)
\begin{equation}
\Omega = -q \left <e|{\bm x}\cdot {\bm \epsilon}_{\rm L}|g \right >  
{\cal E}_{\rm L}\,,
\end{equation}
for a driving laser field ${\bm E}_{\rm L}(t) = {\cal E}_{\rm L} {\bm
\epsilon}_{\rm L} \cos (\omega_{\rm L}t)$ with frequency $\omega_{\rm L}$,
macroscopic classical amplitude ${\cal E}_{\rm L}$ and 
polarization ${\bm \epsilon}_{\rm L}$. Here,
$q=-|q|$ is the elementary charge.
The corresponding coupling constant $g_{\rm L}$ for 
the interaction of a quantized driving laser field with the 
main atomic transition
is defined by
\begin{equation}
g_{\rm L} = 
- q \, \langle g | \bm{\epsilon}_{\rm L} \cdot \bm{x} | e \rangle \,
{\cal E}_{\rm L}^{\rm (\gamma)}\, ,
\end{equation}
where ${\cal E}_{\rm L}^{\rm (\gamma)} = 
\sqrt{\omega_{\rm L}/2V}$ is the electric laser field per photon
and $V$ is the quantization volume. The matching of the electric
field per photon with the corresponding classical macroscopic
electric field ${\cal E}_{\rm L}$ is then given by
\begin{equation}
\label{matching}
2 \: \sqrt{n+1} \: {\cal E}_{\rm L}^{\rm (\gamma)} \: 
\longleftrightarrow \: {\cal E}_{\rm L} \,.
\end{equation}

If $\Omega$
is larger than the natural decay width $\Gamma$ of the 
transition, then the Mollow spectrum approximately consists of
one central and two sideband peaks of Lorentzian shape, which
are located symmetrically around the driving laser field frequency.
The sideband peaks are shifted from the driving field frequency
by the generalized Rabi frequency 
$\Omega_{\textrm R} = \sqrt{\Omega^2 + \Delta^2}$, 
where $\Delta=\omega_{\textrm L} - \omega_{\textrm R}$ 
is the detuning of the laser field frequency $\omega_{\textrm L}$ 
with regard to the atomic transition frequency $\omega_{\textrm R}$. 
The shape of the Mollow spectrum
may easily be explained in terms of the dressed states, which
are defined as the eigenstates of the quantum optical interaction
picture Hamiltonian describing the matter-light interaction.
In transferring to the dressed state picture, the interaction with
the driving laser field is accounted for to all orders.

Thus, when evaluating radiative and relativistic corrections
to the Mollow spectrum, it is natural to start the analysis
from the dressed-state basis as opposed to the
unperturbed atomic bare-state basis. 
In~\cite{JeEvHaKe2003,JeKe2004aop}, it was shown
that this distinction in fact has to be made. It is not 
sufficient to modify the energies (which enter in the
formula for the dressed states) according to the usual
bare-state Lamb shift in order to obtain the correct
result for the corrections to the Mollow spectrum. Instead,
at nonvanishing detuning and nonvanishing Rabi frequency, 
a treatment starting from the dressed-state basis leads
to an additional nontrivial correction term. This term
gives rise to a shift of the Mollow sidebands relative
to the central peak given by
\begin{equation}
\label{deltaomegaplusminus}
\delta \omega_\pm^{(C)} =
\mp {\cal C}\, \frac{\Omega^2}{\sqrt{\Omega^2 + \Delta^2}}\,,
\end{equation}
where
\begin{eqnarray}
{\cal C} &=& \frac{\alpha}{\pi} \, \ln[(Z\alpha)^{-2}] \,
\frac{\left< \bm{p}^2 \right>_g + \left< \bm{p}^2 \right>_e}{m^2}
\end{eqnarray}
is a dimensionless constant. Here, the notation $\langle . \rangle_g$
and $\langle . \rangle_e$ denotes the expectation value evaluated
with the ground or excited atomic state, respectively.

Inspired by the interpretation of the bare Lamb shift correction
in terms of a ``summed'' shift as 
in~\cite{JeEvHaKe2003,JeKe2004aop}, this additional correction 
can be interpreted as a modification
to the Rabi frequency:
\begin{equation}
\label{summedC}
\delta \overline{\omega}^{(C)}_\pm = \pm
\left( \sqrt{\Omega^2 \, (1 - {\cal C})^2 +
\Delta ^2} - \sqrt{\Omega^2 + \Delta^2}\right)\,,
\end{equation}
with $\delta \overline{\omega}^{(C)}_\pm \approx 
\delta\omega_\pm^{(C)}$ because of the 
smallness of the correction. It should be noted that this
interpretation in terms of a summation is not trivial 
and was shown to be valid up to first order in the correction.

In~\cite{JeEvHaKe2003,EvJeKe2004}, the leading relativistic
and radiative corrections up to relative orders $(Z\alpha)^2$
and $\alpha(Z\alpha)^2$, respectively, have been evaluated,
as well as all other relevant correction terms up to the
specified order of approximation. It turns out that all corrections
may be interpreted as either corrections to the Rabi frequency
$\Omega$ or as corrections to the detuning $\Delta$, such that one
can define the fully corrected generalized Rabi frequency
$\Omega^{(j)}_{\mathcal C}$ by 
\begin{equation}
\label{fullOmega}
\Omega^{(j)}_{\mathcal C} =
\sqrt{\Omega^2 \cdot \left(1+{\hat{\Omega}_{\rm rad}^{(j)}}\right)^2 +
\left(\Delta - {\Delta^{(j)}_{\rm rad}}\right)^2}\,.
\end{equation}
Here, $\hat{\Omega}_{\rm rad}^{(j)}$ contains
all corrections to the Rabi frequency, namely
the relativistic and radiative corrections to the transition dipole 
moment, field-configuration dependent corrections, higher-order
corrections to the self-energy, and corrections to the secular 
approximation.
$\Delta^{(j)}_{\rm rad}$
consists of all corrections to the detuning, i.e.
the bare Lamb shift, Bloch-Siegert shifts, and off-resonant
radiative corrections.
The superscript $(j)$ indicates the dependence
of the result on the total angular momentum quantum number.

Equation~(\ref{fullOmega}) summarizes the main result of 
this study: In the presence of driving laser fields, the usual
bare state Lamb shift of the atomic states is augmented
by additional correction terms. These in part depend on the
laser field parameters $\Omega$ and $\Delta$, which span
a two-dimensional parameter manifold determining the
actual value of the dynamical Lamb shift.

A promising candidate for the experiment are
the hydrogen $1S_{\nicefrac{1}{2}}\leftrightarrow 2P_{\nicefrac{1}{2}}$ and 
$1S_{\nicefrac{1}{2}}\leftrightarrow 2P_{\nicefrac{3}{2}}$ 
transitions. 
We consider here
as a specific example the 
$1S_{\nicefrac{1}{2}}\leftrightarrow 2P_{\nicefrac{1}{2}}$ transition
with $\Omega = 1000\cdot \Gamma_{\nicefrac{1}{2}}$ and
$\Delta = 50 \cdot \Gamma_{\nicefrac{1}{2}}$ as the
laser field parameters. The Rabi frequency is shifted
with respect to the leading-order 
expression $\Omega_{\rm R} = \sqrt{\Omega^2 + \Delta^2}$ by
relativistic and radiative corrections as follows,
\begin{equation}
\pm \left (  \Omega^{(\nicefrac{1}{2})}_{\mathcal C} - \Omega_{\rm R} \right ) 
= \pm 738.282(60)\cdot 10^6 \: \rm{Hz}\, .
\end{equation}
This driving laser field parameter set is expected to be within reach 
of improvements of the currently available Lyman-$\alpha$ laser 
sources~\cite{EiWaHa2001} in the next few years.
The corresponding result for the 
$1S_{\nicefrac{1}{2}}\leftrightarrow 2P_{\nicefrac{3}{2}}$ transition 
with $\Omega = 1000\cdot \Gamma_{\nicefrac{3}{2}}$, $\Delta = 50 \cdot \Gamma_{\nicefrac{3}{2}}$
is
\begin{equation}
\pm \left (  \Omega^{(\nicefrac{3}{2})}_{\mathcal C} - \Omega_{\rm R} \right ) 
= \pm 734.871(60)\cdot 10^6 \: \rm{Hz}\, .
\end{equation}
All given uncertainties are due to unknown higher-order
terms~\cite{EvJeKe2004}. 

By a comparison to experimental data,
one may verify the presence of dynamical leading-logarithmic correction
to the dressed-state radiative shift in Eq.~(\ref{deltaomegaplusminus}), 
which cannot
be explained in terms of the bare Lamb shift alone. This allows to 
address
questions related to the physical reality of the dressed states.
On the other hand, the comparison with experimental results could
also be used
to interpret the nature of the evaluated radiative corrections in the sense
of the summation formulas which lead to the interpretation of
the shifts as arising from relativistic and radiative corrections
to the detuning and the Rabi frequency.

%
%
\section{Conclusions}

One of the obvious conclusions to 
be drawn from the recent advances in bound-state
quantum electrodynamics is as follows.
A widespread opinion has been invalidated 
which suggested that two-loop corrections 
to the bound-electron $g$ factor, and two-loop
corrections to the Lamb shift in higher order
might be a severe, if not insurmountable,
obstacle against further theoretical progress.
Quite to the contrary, the recent advances have shown 
that an understanding of these effects is feasible
to a good accuracy. While some important 
contributions remain to be evaluated,
the further program (e.g., the evaluation of 
the remaining high-energy parts)
is clearly defined, and  decisive first 
steps toward a much improved understanding 
have been accomplished.

Further advances are possible both on the 
experimental as well as on the theoretical side: 
concerning the $g$ factor, the recent 
theoretical progress allows for an order-of-magnitude improvement 
of the value for the electron mass based on a potential new measurement alone.
Regarding the Lamb shift in hydrogen and low-$Z$
hydrogenlike systems, one has recently gained 
an improved understanding of the binding corrections of order 
$\alpha^2 \,(Z\alpha)^6\, \ln^j[(Z\alpha)^{-2}]\,
m_{\rm e}\,c^2$ (where $j$ may assume the values $j = 0,\dots,3$).
Recent numerical investigations in the regime 
of intermediate nuclear charge numbers~\cite{YeInSh2004nim}
have also contributed toward an improvement of our understanding.
The extrapolation of the low-$Z$ results
by deferred Pad\'{e} approximants~\cite{Je2003plb} suggests
a rather good general consistency of both approaches,
while some issues regarding the consistency of the
analytic and numerical approaches remain to be 
addressed~\cite{YeInSh2004nim}.

Other progress concerns the modifications of radiative 
corrections in dynamical processed as opposed to 
$S$-matrix energy shifts. The laser-dressed 
Lamb shift~\cite{JeEvHaKe2003,JeKe2004aop,EvJeKe2004,JeEvKe2004}
is a dynamical correction to the dressed-state~\cite{CT1975} 
quasi-energies. The self-energy, in this case,
gives the by far dominant effect. While the bulk of the 
laser-dressed Lamb shift can be understood in terms
of a radiative correction to the detuning, which is 
taken into account in a natural way by evaluating the 
detuning in terms of the observed (low-intensity) 
resonance frequency of the transition, a few tiny 
shifts persist which can only be understood if the 
system is treated in the dressed-state picture right 
from the start. The dynamic Lamb shift is an 
effect which depends on two parameters that determine
the dynamics of the system: (i) the detuning and (ii) the
Rabi frequency. Therefore, 
the dynamical Lamb shift could be mapped out as a function 
of these parameters in a possible experiment.
A rather promising candidate for a possible measurement
would be based on the hydrogen 1S--2P$_{\nicefrac{1}{2}}$ 
transition~\cite{JeEvHaKe2003}. However, a very attractive
alternative would be provided by a forbidden M1 transition~\cite{DrEtAl2003}
in Ar XIV, $2s^2 \,2p\, 
{}^2{\rm P}_{\nicefrac{1}{2}}$--${}^2{\rm P}_{\nicefrac{3}{2}}$,
provided the many-body QED effects can be treated to sufficient
accuracy. The system in question is described very well by 
a two-level formalism, and the resonance line width 
is small.

\section*{Acknowledgments}

The authors acknowledge insightful discussions with 
C. H. Keitel, V. A. Yerokhin and K. Pachucki. 
Helpful conversations with S. G. Karshenboim are also 
gratefully acknowledged.

\end{document}